\documentclass[twocolumn,prl,aps,superscriptaddress,showpacs]{revtex4}

\usepackage{dcolumn}
\usepackage{amsmath}
\usepackage{epsfig}

\newlength{\figwidth}
\newlength{\figwidthb}
\begin{document}

\title{Phonon Softening and Dispersion in $\rm EuTiO_3$}
\author{David S. Ellis}
\email{david_ellis@spring8.or.jp}
\affiliation{Materials Dynamics Laboratory, RIKEN SPring-8 Center, 1-1-1 Kouto, Sayo, Hyogo, 679-5148, Japan}
\author{Hiroshi Uchiyama}
\affiliation{Materials Dynamics Laboratory, RIKEN SPring-8 Center, 1-1-1 Kouto, Sayo, Hyogo, 679-5148, Japan}
\affiliation{Research and Utilization Division, SPring-8/JASRI, Sayo, Hyogo, 679-5198, Japan}
\author{Satoshi Tsutsui}
\affiliation{Research and Utilization Division, SPring-8/JASRI, Sayo, Hyogo, 679-5198, Japan}
\author{Kunihisa Sugimoto}
\affiliation{Research and Utilization Division, SPring-8/JASRI, Sayo, Hyogo, 679-5198, Japan}
\author{Kenichi Kato}
\affiliation{Structural Materials Science Laboratory, RIKEN SPring-8 Center, 1-1-1 Kouto, Sayo, Hyogo, 679-5148, Japan}
\author{Daisuke Ishikawa}
\affiliation{Research and Utilization Division, SPring-8/JASRI, Sayo, Hyogo, 679-5198, Japan}
\author{Alfred Q. R. Baron}
\email{baron@spring8.or.jp}
\affiliation{Materials Dynamics Laboratory, RIKEN SPring-8 Center, 1-1-1 Kouto, Sayo, Hyogo, 679-5148, Japan}
\affiliation{Research and Utilization Division, SPring-8/JASRI, Sayo, Hyogo, 679-5198, Japan}

\date{\today}

\begin{abstract}

We measured phonon dispersion in single crystal EuTiO$_3$ using inelastic x-ray scattering.  A structural transition to an antiferrodistortive phase was found at a critical temperature $T_0$=287$\pm$1 K using powder and single-crystal x-ray diffraction.   Clear softening of the zone boundary \emph{R}-point \textbf{q}=(0.5 0.5 0.5) acoustic phonon shows this to be a displacive transition.  The mode energy plotted against reduced temperature could be seen to nearly overlap that of $\rm SrTiO_3$, suggesting a universal scaling relation.  Phonon dispersion was measured along $\Gamma$-$X$ (0 0 0)$\rightarrow$(0.5 0 0).  Mode eigenvectors were obtained from a shell model consistent with the \textbf{q}-dependence of intensity and energy, which also showed that the dispersion is nominally the same as in $\rm SrTiO_3$ at room temperature, but corrected for mass.  The lowest energy optical mode, determined to be of Slater character, softens approximately linearly with temperature until the 70-100 K range where the softening stops, and at low temperature, the mode disperses linearly near the zone center.


\end{abstract}


\maketitle

Materials in the \emph{A}TiO$_3$ family of perovskites, where \emph{A} = Sr, Ba, Pb and others, have been studied with inelastic neutron scattering \cite{Cowley62,Yamada69,Shirane70,Harada71}, with particular attention to soft phonon modes associated with structural and ferroelectric transitions \cite{Anderson58,Cochran59}.  Although known for nearly 60 years \cite{Brous53}, EuTiO$_3$ had not been included in similar studies, in part due to the neutron-absorbing Eu atoms making such experiments difficult, and to the unavailability of bulk single domain samples necessary to resolve phonons in momentum space.  It is becoming increasingly apparent, however, that EuTiO$_3$ has fascinating properties, owing to the large magnetic spin moments on the Eu$^{2+}$ ion which can couple to the electrical polarization \cite{Katsufuji01,Fennie06,Shvartsman10,Lee10,Mechelen11}.  This makes knowing the lattice dynamical properties of this material all the more imperative, but studies have been largely limited to theoretical efforts \cite{Jiang03,Fennie06,BussmanHolder11,Bettis11,Rushchanskii12} and, on the experimental side, zone-center infra-red reflectivity \cite{Kamba07,Goian09} which probes zero momentum only.  The general consensus is a tendency towards ferroelectricity governed by softening of a zone center $\Gamma$ point optical phonon, which was verified with the infrared reflectivity measurement.

Recently, a peak was detected by Bussmann-Holder \emph{et al}. in the specific heat of a ceramic sample of $\rm EuTiO_3$, at \emph{T}=282 K \cite{BussmanHolder11}.  They hypothesized it originates from an antiferrodistortive transition to a tetragonal I4/\emph{mcm} phase.  This transition features a quasi-rotation of the oxygen octahedra about one of the axes and often a \emph{c/a} lattice constant distortion, and in $\rm SrTiO_3$ exhibits an associated soft phonon at the zone boundary $R$ point in momentum space \cite{Cowley69,Shirane69}.  Since there have been only a handful of materials which clearly showed softening phonons related to this antiferrodistortive transition, and underdamped cases besides $\rm SrTiO_3$ have been quite rare, $\rm EuTiO_3$  therefore potentially represents an important additional window on the dynamics of such soft mode phase transitions.  However, the analysis of  \cite{BussmanHolder11}, which correctly predicted the higher transition temperature for $\rm EuTiO_3$ due to spin-phonon coupling, at the same time suggest that the transitions should have an order-disorder character, as compared to the displacive transition of $\rm SrTiO_3$.  Subsequent x-ray diffraction measurements of $\rm EuTiO_3$ powder samples confirmed the I4/\emph{mcm} low temperature phase for $\rm EuTiO_3$, but only below 235 K \cite{Allieta12}.  The authors of that study argued from pair distribution function analysis that the distortion initially forms in nm-size domains which only order at the lower temperature.  However, a subsequent study \cite{Goian12} observed a transition with powder measurements at around \emph{T}=300K.

To study its dynamics, we grew a bulk single crystal of $\rm EuTiO_3$ for inelastic x-ray scattering, a technique which both bypasses the problem of neutron absorption, and relaxes the need for a large sample.  Our single crystal samples were grown by the traveling float-zone method, using a variant of the method of Katsufuji and Tokura \cite{Katsufuji99}.  Full details of the growth and characterization will be published elsewhere, but a summary is given here.  Feedrods were pressed from a mixture of the component powders Eu$_2$O$_3$ and TiO$_2$, and sintered to form $\rm Eu_2Ti_2O_7$.   Growth was in a reducing 95:5 Ar:H$_2$ gas mixture, using a commercially obtained SrTiO$_3$ seed crystal for the initial run, and iteratively using the top part of previous runs as seeds for the next run.  A SQUID magnometer was used to measure the magnetic susceptibility of the resultant crystal, which showed a peak at \emph{T}$_N$=5.7 K, in accordance with the known N\'{e}el temperature of the antiferromagnetic transition \cite{McGuire66}.  A cleaved surface resulted in the best mosaic, typically 0.01-0.03$^{\circ}$.  We used a (1 0 0) crystal face, which was verified to be consistently single domain over the surface of $\sim$0.5 mm dimension.  At temperatures below \emph{T}=250 K, the crystal became obviously twinned, but the overall mosaic ($\leq$ 0.2$^{\circ}$) and splitting in 2$\theta$ ($|\Delta\textbf{Q}|\sim$0.03 {\AA} ) was deemed sufficiently small to neglect for this study.\\

\begin{figure}
\centering
\epsfig{file=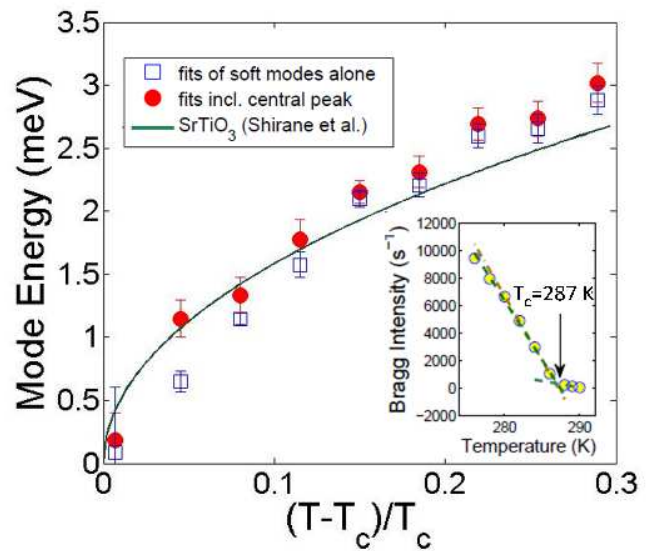,height=3.0in,keepaspectratio, trim=1mm 0 0 0, clip}
\caption{(Color Online)  Energy of the soft mode at \textbf{Q}=(4.5 0.5 1.5), fitting the spectra using two methods.  The empty squares are from fitting a single mode (a stokes/anti-stokes pair), and the filled circles correspond to including a central resolution-limited peak.  As shown in the inset, $T_0$ was determined from the Bragg peak intensity as a function of temperature measured at the \textbf{Q}=(3.5 0.5 0.5) \emph{R}-point.  The continuous curve corresponds to $\rm SrTiO_3$ \cite{Shirane69}, for which $T_0$=108 K.  } \label{fig:T0}
\end{figure}

After observing the twinning, we performed powder diffraction at BL44B2 of SPring-8 \cite{Kato2010}.  The results were broadly consistent with those in \cite{Allieta12}. Our Rietveld refinement \cite{Rodriguez-Carvajal93}, showed a transformation from a cubic phase at room temperature to a tetragonal I4/\emph{mcm} phase below 200 K \cite{fn1}.  Broadening of expected twin peaks was visible below $\sim$240 K as in \cite{Allieta12}, however weak $R$-point (1.5 0.5 0.5) intensity was visible in our powder data below $\sim$280 K, whereas it was only seen at $\sim$200 K in \cite{Allieta12}.  We also investigated the $R$-point (3.5 0.5 0.5) reflection using single crystal diffraction.  The intensity had a clear onset at 287$\pm$1 K (see inset to fig.~\ref{fig:T0}), which is within a few degrees of the previous calorimetric data \cite{BussmanHolder11}. The angular width of the diffracted beam was largely temperature independent, varying from 0.23 to 0.33 mrad depending on location on the sample, at both 280 K and 220 K.  From this we estimate a minimum correlation length  $>$0.1 $\mu$m.  This suggests that the domain sizes are large, which does not appear consistent with disorder of the oxygen octahedra rotation angles suggested in \cite{Allieta12}. \\

\begin{figure}
\centering
\epsfig{file=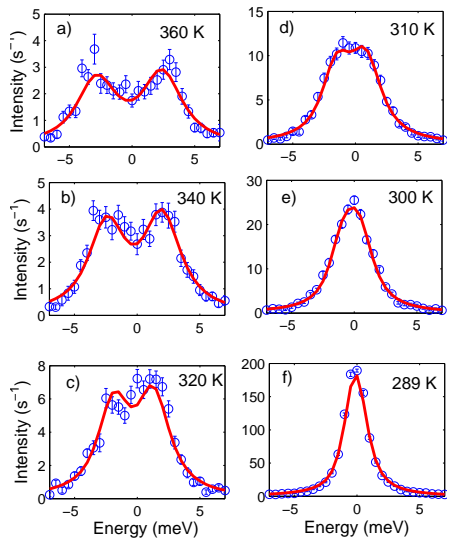,height=2.8in,keepaspectratio, trim=1mm 0 0 0, clip}
\caption{(Color Online)  (a)-(f) Selected phonon spectra at the $R$-point (\textbf{Q}=(4.5 0.5 1.5), with resolution $\delta$\textbf{Q}=(0.03 0.02 0.03)) as a function of temperature as indicated.  The solid lines are fits to a phonon mode (with no central component) as described in the text.} \label{fig:Rpoint}
\end{figure}

Inelastic x-ray scattering measurements were done using BL35XU at SPring-8 \cite{Baron2000}, using a 21.7 keV incident energy with typically 1.5 meV resolution and momentum resolution of 0.03, in units of reciprocal lattice.  The \emph{R}-point spectrum at \textbf{Q}=(4.5 0.5 1.5) was measured between \emph{T}=370 K and \emph{T}=289 K at intervals of 10 K.  Selected spectra are shown in Fig.~\ref{fig:Rpoint}(a)-(f), which clearly show a softening mode.  The data were fitted to two Lorentzians, whose peak intensities are constrained by detailed balance, and convolved with the resolution function.  The fitted energies are plotted against reduced temperature \emph{t}=$\frac{T-T_0}{T_0}$ in Fig.~\ref{fig:T0}.   In fitting the spectra we first neglected, and then considered, a possible ``central mode'' at zero energy \cite{Riste71,Shapiro72}.  The mode energies resulting from fits which include it as a resolution-limited peak of freely varying intensity, are also plotted in Fig.~\ref{fig:T0}.  Near the transition, the energy derived without including a central peak is linear with temperature, and fitting to a line gives $T_0$=287 K.  This is consistent with value of $T_0$=287$\pm$1 K as determined from the Bragg intensity shown in the inset of fig.~\ref{fig:T0}.  For a comparison with $\rm SrTiO_3$, we plot its soft mode energy in fig.~\ref{fig:T0}  using the empirical formula of Shirane and Yamada \cite{Shirane69}, who determined $T_0$=108 K.  With the second fitting method, the approach to $T_0$ in the two materials becomes identical within error bar for $\frac{T-T_0}{T_0}$$\leq$0.1.  Previously, M$\rm \ddot{u}$ller and Berlinger had found a universality in the critical exponent of the order parameter below the transition temperature, by comparing $\rm SrTiO_3$ to $\rm LaAlO_3$ \cite{Muller71}, but the present result, assuming the plausible existence of a central component, suggests a further universality in the soft mode energy vs. reduced temperature.  That the \emph{absolute} phonon energies would line up may stem from a similarity between the two systems.\\

\begin{figure}[ht!]
\centering
\epsfig{file=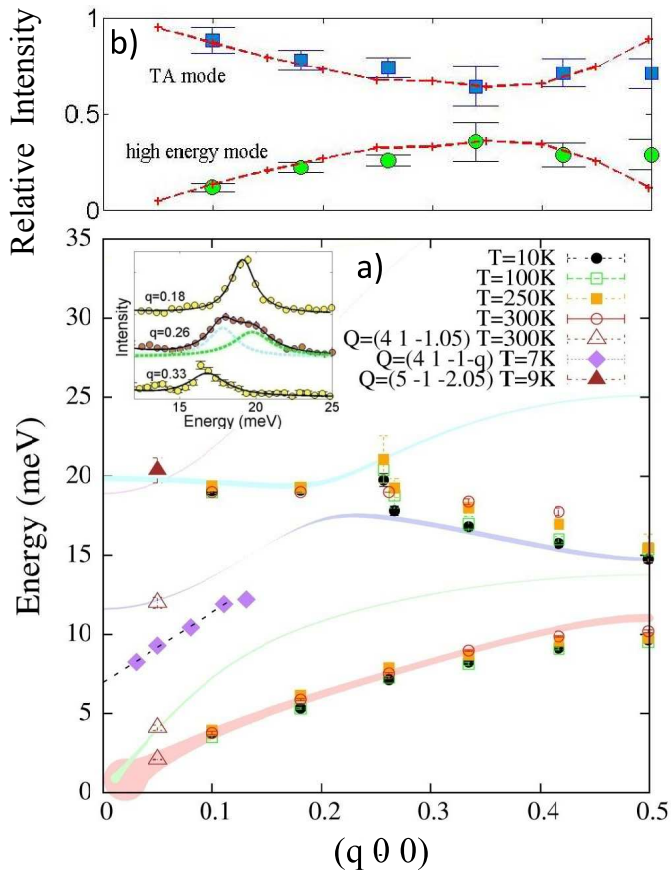,totalheight=3.7in,angle=-90,keepaspectratio, trim=1mm 0 0 0, clip}
\caption{(Color Online) (a) (bottom panel) Phonon energy vs. momentum \emph{q} along the $\Gamma$-$X$ direction.  The dispersion away from the $\Gamma$-point (\emph{q}$\geq$0.1) was for (\emph{H} \emph{K} \emph{L})=(4, 0, 1-\emph{q}).  For the near-$\Gamma$ dispersion,  zones (4 1 -1) (empty triangles and diamonds) and (5 -1 -2) (filled triangles) were used as indicated in the legend.  The smooth curves are model calculations, with the thickness of the curve proportional to the square root of the intensity in the (4 0 1) zone.  The inset shows the spectra at the anti-crossing near 20 meV and \emph{q}=0.25 at \emph{T}=10 K.  For clarity, the two distinct mode energies at \emph{q}=0.26 have been slightly offset along the \emph{q} axis. (b) (top panel) Relative intensities of the transverse acoustic and high-energy modes along (4, 0, 1-\emph{q}), comparing the experimental values obtained from fits to the data for $T$=250 K, and calculated values from the shell model, which are the dashed lines.} \label{fig:disp}

\end{figure}

Energy spectra as a function of momentum and temperature were measured along the $\Gamma$-$X$ direction.  The phonon dispersion curves, obtained from fits to our data, are shown in fig.~\ref{fig:disp}(a).  Some typical spectra can be found in the inset.  We also overlay the dispersion curve calculated using a shell model allowing for displacement of the atom's core and outer electron shell \cite{Dick58}, implemented with the OpenPhonon code \cite{Openphonon}.  As shown in Fig.~\ref{fig:disp}(a), the theoretical intensities of each mode, as represented by the thicknesses of curves, can vary greatly within the zone, to the extent that two modes may be mistaken for one when the intensity of one falls and the other rises.  For example, modes in the 15-20 meV range which broaden at \emph{q}=0.26 are shown in the inset of Fig.~\ref{fig:disp}(a), and the broadening can be qualitatively explained by an anticrossing of two modes, predicted in the model.  As further verification of the model, we plot in fig.~\ref{fig:disp}(b) the \emph{q}-dependence of the relative integrated mode intensities at $T$=250 K, of the transverse acoustic mode and high-energy mode(s) that can be found in the 14-21 meV range, and normalize by the sum of the two intensities.  By comparing the relative intensities of the modes, systematic error in the intensity that may arise from the beam hitting different parts of the sample as \textbf{q} is changed is canceled out. Also plotted is the corresponding calculated intensity fraction from the shell model for modes within the prescribed range, which is found to be in excellent agreement.   We should note that our model parameters were originally optimized using $\rm SrTiO_3$ dispersion data along the $\Gamma$-$X$ and $\Gamma$-$M$ directions \cite{Cowley64, Stirling72}.  The only subsequent modification of the parameters was to change the mass (and atomic scattering factors affecting intensity) from Sr to Eu , whose most noticeable effect was a significant lowering of the acoustic mode frequencies.

The calculated eigenvectors describing the atomic motion are shown in table 1.  The first two entries are for the lowest two optical modes calculated for \emph{q}=0, TO$_1$ and TO$_2$.  The last entry in the table is the lowest energy mode at the $R$ point. It consists of oxygen motions in the plane of the faces of the conventional cubic cell, and is a linear combination of degenerate modes of the form $O_{ij}$=$-O_{ji}$ where $O_{ij}$ is the displacement of the oxygen on the $i$-facing plane in the $\rm \emph{j}^{th}$ direction.  Such modes condense into the I4/\emph{mcm} phase and correspond to the $R$-point soft mode \cite{Shirane69}.  Table 1 confirms that the TO$_1$ mode is of ``Slater" chatacter, similar to the soft mode in $\rm SrTiO_3$ \cite{Cowley64,Harada70}, while TO$_2$ has Eu and Ti motions 180$^\circ$ out of phase.  Looking at the temperature dependence of the modes in fig.~\ref{fig:disp}, the energy of the TO$_1$ mode is generally the most sensitive to temperature in the zone.\\

\begin{table}

\caption {Table of calculated cubic-phase eigenvectors for the modes TO$_1$ at around 11.7 meV and TO$_2$ at 20 meV at the $\Gamma$-point, and a \textbf{q}=(0.5 0.5 0.5) $R$-point mode, denoted TA$_R$.  The columns are the atom, the equilibrium position, followed by the displacements for each mode in units of lattice constant.}

\begin{ruledtabular}

\begin{tabular}{ccccc}

Atom & Position  & TO$_1$ & TO$_2$  & TA$_R$ \\
\hline

Eu & ($\frac{1}{2}$,$\frac{1}{2}$,$\frac{1}{2}$) & (-0.007, 0, 0) & (-0.011, 0, 0) & (0, 0, 0)\\

Ti  & (0,0,0) & (-0.011,0,0) & (0.03,0,0) & (0,0,0) \\

O & ($\frac{1}{2}$,0,0) &  (0.018,0,0) & (0.016,0,0) & (0,-0.036,0.018)\\

O & (0,$\frac{1}{2}$,0) & (0.038,0,0) & (0,0,0) & (0.036,0,-0.018)\\

O & (0,0,$\frac{1}{2}$) & (0.038,0,0) & (0,0,0) & (-0.018,0.018,0)\\

\end{tabular}

\end{ruledtabular}

\end{table}

\begin{figure}
\setlength{\belowcaptionskip}{-15pt}
\centering
\epsfig{file=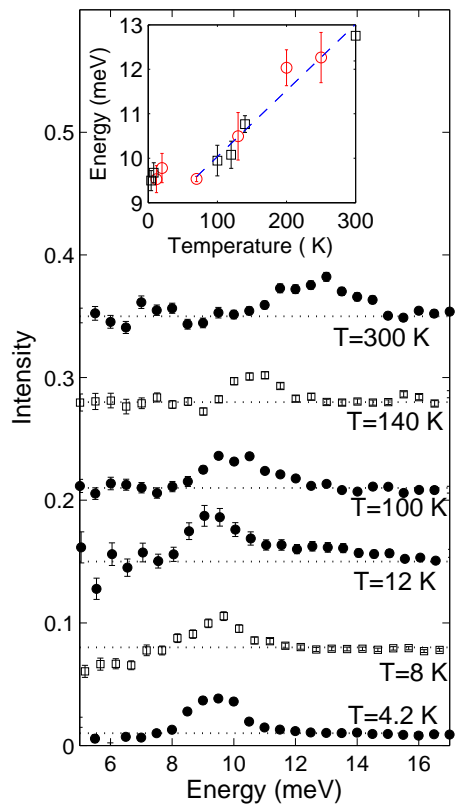,height=4.2in,keepaspectratio, trim=1mm 0 0 0, clip}
\caption{(Color Online) Selected phonon spectra near the gamma point at \emph{q}=0.05.  Energy offsets were accounted for through fits of elastic intensity and acoustic modes on either side of $\hbar\omega$=0 where applicable.  A smooth background was subtracted from the curves, whose intensities are corrected for Bose factor.  The intensities were further normalized by the Bose-factor corrected transverse acoustic mode intensity. The Q position used was (5 -1 -2.05) except for T=12 K which was taken at Q=(4 1 -1.05).  Fitted energy vs. temperature is plotted in the inset for the (4 1 -1.05) and (5 -1 -2.05) zones (circles and squares, respectively).} \label{fig:gammapoint}
\end{figure}

To investigate the temperature dependence of the TO$_1$ mode near the zone center, we changed to the (4 1 -1) and (5 -1 -2) zones where our calculations predict a stronger intensity of that mode at the $\Gamma$-point.  We note that even in these favorable zones, the mode is relatively weak compared to background coming from the tail of the acoustic modes, at high temperatures.  Background-subtracted spectra, normalized for Bose factor and further by acoustic mode peak, are plotted in fig.~\ref{fig:gammapoint} for constant $q$=0.05 at selected temperatures.  This latter normalization was to use the acoustic intensity as a relatively stable reference against different sample positions at different temperatures, which could affect intensity.  The ferroelectric mode energy vs. temperature, measured over two zones, is plotted in the inset, including energies from additional spectra not plotted in the main figure.   There is a general lowering of energy as the temperature is lowered,  which is same order of magnitude previously observed in infrared reflectivity \cite{Kamba07,Goian09} (we estimate a $\sim$4.5 meV shift by extrapolation to \emph{q}=0, based on the dispersion in Fig.~\ref{fig:disp}), and the trend is approximately linear down to about 70-100 K after which the softening stops.  The low temperature dispersion near \emph{q}=0  is plotted in fig.~\ref{fig:disp} as diamonds.  Although high elastic background near zone center prevents measurement at exactly \emph{q}=0, the dispersion down to at least \emph{q}=0.02 is somewhat more linear than the room-temperature model would suggest for that mode in Fig.~\ref{fig:disp}(a), where the dispersion instead flattens out as \emph{q} is lowered.  This linear dispersion characteristic of a soft mode \cite{Yamada69}.\\

We observe no resolvable energy change below the magnetic transition $T_N$=5.7 K.  This result may initially be surprising (not to mention disappointing), given the dramatic drop in dielectric constant measured in \cite{Katsufuji01}, however we note that the magnitude of the dielectric constant drop within 1 K of the transition is less than 0.3 \%.  A recent study of the optic mode using infrared reflectivity on thin $\rm EuTiO3$ films \cite{Kamba12} films did in fact show an increase in phonon frequency below $T_N$, but in the vicinity of the transition, the change is less than 1\%, which is indeed less than the present experiment could resolve for this mode. \\

In summary, we prepared a single domain crystal of $\rm EuTiO_3$ and observed it to undergo a transition into a structurally long-range ordered phase at \emph{T}=287$\pm$1 K, consistent with the I4/\emph{mcm} structure from our powder refinements.  The combined phonon and super-lattice Bragg peak intensity observations, which have consistent \emph{T}$_0$, clearly prove a displacive structural transition.  In this respect, the picture is considerably simpler than other recent diffraction studies in the literature \cite{Allieta12,Goian12,Kim12}, where in some cases time-dependent incommensurabilities were observed.  Sensitivity to sample preparation conditions may be a factor.  We note that our data was taken over multiple beamtimes, and the phonons were basically consistent throughout.   Near the structural transition, the $R$-point soft phonon exhibits almost identical critical behavior to $\rm SrTiO_3$, with the same phonon energy vs. reduced temperature curve, suggesting a universal scaling relation for the soft mode energy.  The lowest energy optical mode exhibits soft mode behavior consistent with displacive-type dynamics. A mass-corrected $\rm SrTiO_3$-based shell model can fairly accurately describe the frequencies and intensities of $\rm EuTiO_3$ phonons, which further emphasizes the close similarity in the two systems.    \\

The authors thank Yoshikazu Tanaka for help with the use of the x-ray generator facilities for preliminary characterization of our materials and samples.  We also gratefully acknowledge Masayuki Udagawa for use of He transfer tube, and Tatsuo Fukuda for helpful discussions about the shell model, and Seigo Yamamoto for his invaluable help in preparing for the powder measurements.  We thank the Japan Synchrotron Radiation Research Institute (JASRI) for granting us beam time at BL35XU in SPring-8 under proposal no.'s 2010B1527, 2011A1271, 2011B1590, 2012A1362, 2012A1818, and RIKEN for granting us beamtime at the RIKEN Materials Science Beamline (BL44B2) in SPring-8 under proposal no. 2011-2881.



\begin{thebibliography}{56}
\expandafter\ifx\csname natexlab\endcsname\relax\def\natexlab#1{#1}\fi
\expandafter\ifx\csname bibnamefont\endcsname\relax
  \def\bibnamefont#1{#1}\fi
\expandafter\ifx\csname bibfnamefont\endcsname\relax
  \def\bibfnamefont#1{#1}\fi
\expandafter\ifx\csname citenamefont\endcsname\relax
  \def\citenamefont#1{#1}\fi
\expandafter\ifx\csname url\endcsname\relax
  \def\url#1{\texttt{#1}}\fi
\expandafter\ifx\csname urlprefix\endcsname\relax\def\urlprefix{URL }\fi
\providecommand{\bibinfo}[2]{#2}
\providecommand{\eprint}[2][]{\url{#2}}


\bibitem[{\citenamefont{Yamada and Shirane}(1969)}]{Yamada69}
\bibinfo{author}{\bibfnamefont{Yasusada} \bibnamefont{Yamada}} \bibnamefont{and}
\bibinfo{author}{\bibfnamefont{Gen} \bibnamefont{Shirane}},
\bibinfo{journal}{Journal of the Phys. Soc. of Japan} \textbf{\bibinfo{volume}{26}},
\bibinfo{pages}{396} (\bibinfo{year}{1969}).

\bibitem[{\citenamefont{Shirane}(1970)}]{Shirane70}
\bibinfo{author}{\bibfnamefont{G.} \bibnamefont{Shirane}},
\bibinfo{author}{\bibfnamefont{J.~D.} \bibnamefont{Axe}},
\bibinfo{author}{\bibfnamefont{J.} \bibnamefont{Harada}} \bibnamefont{and}
\bibinfo{author}{\bibfnamefont{J.~.P} \bibnamefont{Remeika}},
\bibinfo{journal}{\prb} \textbf{\bibinfo{volume}{2}},
\bibinfo{pages}{155} (\bibinfo{year}{1970}).


\bibitem[{\citenamefont{Harada}(1971)}]{Harada71}
\bibinfo{author}{\bibfnamefont{J.} \bibnamefont{Harada}},
\bibinfo{author}{\bibfnamefont{J.~D.} \bibnamefont{Axe}} \bibnamefont{and}
\bibinfo{author}{\bibfnamefont{G.} \bibnamefont{Shirane}},
\bibinfo{journal}{\prb} \textbf{\bibinfo{volume}{4}},
\bibinfo{pages}{155} (\bibinfo{year}{1971}).

\bibitem[{\citenamefont{Cowley}(1962)}]{Cowley62}
\bibinfo{author}{\bibfnamefont{R.~A.} \bibnamefont{Cowley}},
\bibinfo{journal}{\prl} \textbf{\bibinfo{volume}{9}},
\bibinfo{pages}{159} (\bibinfo{year}{1962}).

\bibitem[{\citenamefont{Anderson}(1958)}]{Anderson58}
\bibinfo{author}{\bibfnamefont{P.~W.} \bibnamefont{Anderson}},
\bibinfo{journal}{Academy of Science, U.S.S.R. Moscow},
\bibinfo{pages}{p. 290} (\bibinfo{year}{1958}).

\bibitem[{\citenamefont{Chochran}(1959)}]{Cochran59}
\bibinfo{author}{\bibfnamefont{W} \bibnamefont{Cochran}},
\bibinfo{journal}{\prl} \textbf{\bibinfo{volume}{3}},
\bibinfo{pages}{412} (\bibinfo{year}{1959}).

\bibitem[{\citenamefont{Brous et al}(1953)}]{Brous53}
\bibinfo{author}{\bibfnamefont{J.} \bibnamefont{Brous}},
\bibinfo{author}{\bibfnamefont{I.} \bibnamefont{Fankuchen}} \bibnamefont{and}
\bibinfo{author}{\bibfnamefont{E.} \bibnamefont{Banks}},
\bibinfo{journal}{Acta Cryst.} \textbf{\bibinfo{volume}{6}},
\bibinfo{pages}{67} (\bibinfo{year}{1953}).

\bibitem[{\citenamefont{Katsufuji and Takagi}(2001)}]{Katsufuji01}
\bibinfo{author}{\bibfnamefont{T.} \bibnamefont{Katsufuji}} \bibnamefont{and}
  \bibinfo{author}{\bibfnamefont{H.} \bibnamefont{Takagi}},
  \bibinfo{journal}{\prb} \textbf{\bibinfo{volume}{64}},
  \bibinfo{pages}{054415} (\bibinfo{year}{2001}).

  \bibitem[{\citenamefont{Fennie and Rabe}(2006)}]{Fennie06}
\bibinfo{author}{\bibfnamefont{C.~J} \bibnamefont{Fennie}} \bibnamefont{and}
\bibinfo{author}{\bibfnamefont{K.~M} \bibnamefont{Rabe}},
\bibinfo{journal}{\prl} \textbf{\bibinfo{volume}{97}},
\bibinfo{pages}{267602} (\bibinfo{year}{2006}).

\bibitem[{\citenamefont{Shvartsman et al}(2010)}]{Shvartsman10}
\bibinfo{author}{\bibfnamefont{V.~V.} \bibnamefont{Shvartsman}},
\bibinfo{author}{\bibfnamefont{P.} \bibnamefont{Borisov}},
\bibinfo{author}{\bibfnamefont{W.} \bibnamefont{Kleemann}},
\bibinfo{author}{\bibfnamefont{S.} \bibnamefont{Kamba}} \bibnamefont{and}
\bibinfo{author}{\bibfnamefont{T.} \bibnamefont{Katsufuji}},
\bibinfo{journal}{\prb} \textbf{\bibinfo{volume}{81}},
\bibinfo{pages}{064426} (\bibinfo{year}{2010}).

\bibitem[{\citenamefont{Lee et al}(2010)}]{Lee10}
\bibinfo{author}{\bibfnamefont{June Hyuk} \bibnamefont{Lee}},
\bibinfo{author}{\bibfnamefont{Lei} \bibnamefont{Fang}},
\bibinfo{author}{\bibfnamefont{Eftihia} \bibnamefont{Vlahos}},
\bibinfo{author}{\bibfnamefont{Xianglin} \bibnamefont{Ke}},
\bibinfo{author}{\bibfnamefont{Young Woo} \bibnamefont{Jung}},
\bibinfo{author}{\bibfnamefont{Lena Fitting} \bibnamefont{Kourkoutis}},
\bibinfo{author}{\bibfnamefont{Jong-Woo} \bibnamefont{Kim}},
\bibinfo{author}{\bibfnamefont{Philip J.} \bibnamefont{Ryan}},
\bibinfo{author}{\bibfnamefont{Tassilo} \bibnamefont{Heeg}},
\bibinfo{author}{\bibfnamefont{Martin} \bibnamefont{Roeckerath}},
\bibinfo{author}{\bibfnamefont{Veronica} \bibnamefont{Goian}},
\bibinfo{author}{\bibfnamefont{Margitta} \bibnamefont{Bernhagen}},
\bibinfo{author}{\bibfnamefont{Reinhard} \bibnamefont{Uecker}},
\bibinfo{author}{\bibfnamefont{P. Chris} \bibnamefont{Hammel}},
\bibinfo{author}{\bibfnamefont{Karin M.} \bibnamefont{Rabe}},
\bibinfo{author}{\bibfnamefont{Stanislav} \bibnamefont{Kamba}},
\bibinfo{author}{\bibfnamefont{J\"{u}rgen} \bibnamefont{Schubert}},
\bibinfo{author}{\bibfnamefont{John W.} \bibnamefont{Freeland}},
\bibinfo{author}{\bibfnamefont{David A.} \bibnamefont{Muller}},
\bibinfo{author}{\bibfnamefont{Craig J.} \bibnamefont{Fennie}},
\bibinfo{author}{\bibfnamefont{Peter} \bibnamefont{Schiffer}},
\bibinfo{author}{\bibfnamefont{Venkatraman} \bibnamefont{Gopalan}},
\bibinfo{author}{\bibfnamefont{Ezekiel} \bibnamefont{Johnston-Halperin}} \bibnamefont{and}
\bibinfo{author}{\bibfnamefont{Darrell G.} \bibnamefont{Schlom}},
\bibinfo{journal}{Nature} \textbf{\bibinfo{volume}{466}},
\bibinfo{pages}{954} (\bibinfo{year}{2006}).

\bibitem[{\citenamefont{Mechelen et al}(2011)}]{Mechelen11}
\bibinfo{author}{\bibfnamefont{J.~L.~M.~van} \bibnamefont{Mechelen}},
\bibinfo{author}{\bibfnamefont{D.~van~der} \bibnamefont{Marel}} \bibnamefont{and}
\bibinfo{author}{\bibfnamefont{I.} \bibnamefont{Crassee}},
\bibinfo{author}{\bibfnamefont{T.} \bibnamefont{Kolodiazhnyi}},
\bibinfo{journal}{\prl} \textbf{\bibinfo{volume}{106}},
\bibinfo{pages}{217601} (\bibinfo{year}{2011}).

\bibitem[{\citenamefont{Jiang and Wu}(2003)}]{Jiang03}
\bibinfo{author}{\bibfnamefont{Qing} \bibnamefont{Jiang}} \bibnamefont{and}
\bibinfo{author}{\bibfnamefont{Hua} \bibnamefont{Wu}},
\bibinfo{journal}{J. Appl. Phys.} \textbf{\bibinfo{volume}{93}},
\bibinfo{pages}{2121} (\bibinfo{year}{2003}).

\bibitem[{\citenamefont{Bussmann-Holder et al}(2011)}]{BussmanHolder11}
\bibinfo{author}{\bibfnamefont{A.} \bibnamefont{Bussmann-Holder}},
\bibinfo{author}{\bibfnamefont{J.} \bibnamefont{K\"{o}hler}},
\bibinfo{author}{\bibfnamefont{R.~K.} \bibnamefont{Kremer}} \bibnamefont{and}
\bibinfo{author}{\bibfnamefont{J.~M.} \bibnamefont{Law}},
\bibinfo{journal}{\prb} \textbf{\bibinfo{volume}{83}},
\bibinfo{pages}{212102} (\bibinfo{year}{2011}).

\bibitem[{\citenamefont{Bettis}(2011)}]{Bettis11}
\bibinfo{author}{\bibfnamefont{Jerry~L.} \bibnamefont{Bettis}},
\bibinfo{author}{\bibfnamefont{Myung-Hwan} \bibnamefont{Whangbo}},
\bibinfo{author}{\bibfnamefont{J\"{u}rgen} \bibnamefont{K\"{o}hler}},
\bibinfo{author}{\bibfnamefont{Annette} \bibnamefont{Bussmann-Holder}} \bibnamefont{and}
\bibinfo{author}{\bibfnamefont{A.~R.} \bibnamefont{Bishop}},
\bibinfo{journal}{\prb} \textbf{\bibinfo{volume}{84}},
\bibinfo{pages}{184114} (\bibinfo{year}{2011}).

\bibitem[{\citenamefont{Rushchanskii}(2012)}]{Rushchanskii12}
\bibinfo{author}{\bibfnamefont{Konstantin~Z.} \bibnamefont{Rushchanskii}},
\bibinfo{author}{\bibfnamefont{Nicola~A.} \bibnamefont{Spaldin}} \bibnamefont{and}
\bibinfo{author}{\bibfnamefont{Marjana} \bibnamefont{Le\v{z}ai\'{c}}},
\bibinfo{journal}{\prb} \textbf{\bibinfo{volume}{85}},
\bibinfo{pages}{104109} (\bibinfo{year}{2012}).

\bibitem[{\citenamefont{Kamba et al}(2007)}]{Kamba07}
\bibinfo{author}{\bibfnamefont{S.} \bibnamefont{Kamba}},
\bibinfo{author}{\bibfnamefont{D.} \bibnamefont{Nuzhnyy}},
\bibinfo{author}{\bibfnamefont{P.} \bibnamefont{Van\v{e}k}},
\bibinfo{author}{\bibfnamefont{M.} \bibnamefont{Savinov}},
\bibinfo{author}{\bibfnamefont{K.} \bibnamefont{Kn\'{i}\v{z}ek}},
\bibinfo{author}{\bibfnamefont{Z.} \bibnamefont{Shen}},
\bibinfo{author}{\bibfnamefont{E.} \bibnamefont{\v{S}antav\'{a}}},
\bibinfo{author}{\bibfnamefont{K.} \bibnamefont{Maca}},
\bibinfo{author}{\bibfnamefont{M.} \bibnamefont{Sadowski}} \bibnamefont{and}
\bibinfo{author}{\bibfnamefont{J.} \bibnamefont{Petzelt}},
\bibinfo{journal}{European Physics Letters} \textbf{\bibinfo{volume}{80}},
\bibinfo{pages}{27002} (\bibinfo{year}{2007}).


\bibitem[{\citenamefont{Goian et al}(2009)}]{Goian09}
\bibinfo{author}{\bibfnamefont{V.} \bibnamefont{Goian}},
\bibinfo{author}{\bibfnamefont{S.} \bibnamefont{Kamba}},
\bibinfo{author}{\bibfnamefont{J.} \bibnamefont{Hlinka}},
\bibinfo{author}{\bibfnamefont{P.} \bibnamefont{Van\v{e}k}},
\bibinfo{author}{\bibfnamefont{A.~A.} \bibnamefont{Belik}},
\bibinfo{author}{\bibfnamefont{T.} \bibnamefont{Kolodiazhnyi}} \bibnamefont{and}
\bibinfo{author}{\bibfnamefont{J.} \bibnamefont{Petzelt}},
\bibinfo{journal}{Eur. Phys. J. B} \textbf{\bibinfo{volume}{71}},
\bibinfo{pages}{429} (\bibinfo{year}{2009}).

\bibitem[{\citenamefont{Shirane and Yamada}(1969)}]{Shirane69}
\bibinfo{author}{\bibfnamefont{G.} \bibnamefont{Shirane}} \bibnamefont{and}
\bibinfo{author}{\bibfnamefont{Y.} \bibnamefont{Yamada}},
\bibinfo{journal}{Phys. Rev.} \textbf{\bibinfo{volume}{177}},
\bibinfo{pages}{858} (\bibinfo{year}{1969}).

\bibitem[{\citenamefont{Cowley et al.}(1969)}]{Cowley69}
\bibinfo{author}{\bibfnamefont{R.~A.} \bibnamefont{Cowley}},
\bibinfo{author}{\bibfnamefont{W.~J.~L.} \bibnamefont{Buyers}} \bibnamefont{and}
\bibinfo{author}{\bibfnamefont{G.} \bibnamefont{Dolling}},
\bibinfo{journal}{Solid St. Commun.} \textbf{\bibinfo{volume}{7}},
\bibinfo{pages}{181} (\bibinfo{year}{1969}).


\bibitem[{\citenamefont{Allieta et al}(2012)}]{Allieta12}
\bibinfo{author}{\bibfnamefont{Mattia} \bibnamefont{Allieta}},
\bibinfo{author}{\bibfnamefont{Marco} \bibnamefont{Scavini}},
\bibinfo{author}{\bibfnamefont{Leszek~J.} \bibnamefont{Spalek}},
\bibinfo{author}{\bibfnamefont{Valerio} \bibnamefont{Scagnoli}},
\bibinfo{author}{\bibfnamefont{Helen~C.} \bibnamefont{Walker}},
\bibinfo{author}{\bibfnamefont{Christos} \bibnamefont{Panagopoulos}},
\bibinfo{author}{\bibfnamefont{Siddharth~S.} \bibnamefont{Saxena}},
\bibinfo{author}{\bibfnamefont{Takuro} \bibnamefont{Katsufuji}} \bibnamefont{and}
\bibinfo{author}{\bibfnamefont{Claudio} \bibnamefont{Mazzoli}},
\bibinfo{journal}{\prb} \textbf{\bibinfo{volume}{85}},
\bibinfo{pages}{184107} (\bibinfo{year}{2012}).

\bibitem[{\citenamefont{Katsufuji and Tokura}(1999)}]{Katsufuji99}
\bibinfo{author}{\bibfnamefont{T.} \bibnamefont{Katsufuji}} and
\bibinfo{author}{\bibfnamefont{Y.} \bibnamefont{Tokura}},
\bibinfo{journal}{\prb} \textbf{\bibinfo{volume}{60}},
\bibinfo{pages}{R15021} (\bibinfo{year}{1999}).

\bibitem[{\citenamefont{McGuire et al}(1966)}]{McGuire66}
\bibinfo{author}{\bibfnamefont{T.~R.} \bibnamefont{McGuire}},
\bibinfo{author}{\bibfnamefont{M.~W.} \bibnamefont{Shafer}},
\bibinfo{author}{\bibfnamefont{R.~J.} \bibnamefont{Joenk}},
\bibinfo{author}{\bibfnamefont{H.~A.} \bibnamefont{Alperin}} \bibnamefont{and}
\bibinfo{author}{\bibfnamefont{S.~J.} \bibnamefont{Pickart}},
\bibinfo{journal}{J. Applied Phys.} \textbf{\bibinfo{volume}{37}},
\bibinfo{pages}{981} (\bibinfo{year}{1966}).

\bibitem[{\citenamefont{Kato et al}(2010)}]{Kato2010}
\bibinfo{author}{\bibfnamefont{K.} \bibnamefont{Kato}},
\bibinfo{author}{\bibfnamefont{R.} \bibnamefont{Hirose}},
\bibinfo{author}{\bibfnamefont{M.} \bibnamefont{Takemoto}},
\bibinfo{author}{\bibfnamefont{S.} \bibnamefont{Ha}},
\bibinfo{author}{\bibfnamefont{J.} \bibnamefont{Kim}},
\bibinfo{author}{\bibfnamefont{M.} \bibnamefont{Higuchi}},
\bibinfo{author}{\bibfnamefont{R.} \bibnamefont{Matsuda}},
\bibinfo{author}{\bibfnamefont{S.} \bibnamefont{Kitagawa}} \bibnamefont{and}
\bibinfo{author}{\bibfnamefont{M.} \bibnamefont{Takata}},
\bibinfo{journal}{AIP Conference Proceedings} \textbf{\bibinfo{volume}{1234}},
\bibinfo{pages}{2010} (\bibinfo{year}{875}).

\bibitem[{\citenamefont{FullProf}(1999)}]{Rodriguez-Carvajal93}
\bibinfo{author}{ \bibnamefont{Using the FullProf program; see}}
\bibinfo{author}{\bibfnamefont{J.} \bibnamefont{Rodriguez-Carvajal}},
\bibinfo{journal}{Physica B} \textbf{\bibinfo{volume}{192}},
\bibinfo{pages}{55} (\bibinfo{year}{1993}).

\bibitem[{tetr()}]{fn1}
{The region between 200 K and 287 K could be refined well with either the cubic or I4/\emph{mcm} structure.  One recent paper \cite{Goian12} has suggested the I4/\emph{mcm} structure for this region.}

\bibitem[{\citenamefont{Goian et al}(2012)}]{Goian12}
\bibinfo{author}{\bibfnamefont{V.} \bibnamefont{Goian}},
\bibinfo{author}{\bibfnamefont{S.} \bibnamefont{Kamba}},
\bibinfo{author}{\bibfnamefont{O.} \bibnamefont{Pacherov$\rm \acute{a}$}},
\bibinfo{author}{\bibfnamefont{J.} \bibnamefont{Drahokoupil}},
\bibinfo{author}{\bibfnamefont{L.} \bibnamefont{Palatinus}},
\bibinfo{author}{\bibfnamefont{M.} \bibnamefont{Du$\rm \check{s}$ek}},
\bibinfo{author}{\bibfnamefont{J.} \bibnamefont{Rohl$\rm \acute{i}$$\rm \check{c}$ek}},
\bibinfo{author}{\bibfnamefont{M.} \bibnamefont{Savinov}},
\bibinfo{author}{\bibfnamefont{F.} \bibnamefont{Laufek}},
\bibinfo{author}{\bibfnamefont{W.} \bibnamefont{Schranz}},
\bibinfo{author}{\bibfnamefont{A.} \bibnamefont{Fuith}},
\bibinfo{author}{\bibfnamefont{M.} \bibnamefont{Kachl$\rm \acute{i}$k}},
\bibinfo{author}{\bibfnamefont{K.} \bibnamefont{Maca}},
\bibinfo{author}{\bibfnamefont{A.} \bibnamefont{Shkabko}},
\bibinfo{author}{\bibfnamefont{L.} \bibnamefont{Sagarna}},
\bibinfo{author}{\bibfnamefont{A.} \bibnamefont{Weidenkaff}}\bibnamefont{and}
\bibinfo{author}{\bibfnamefont{A.~A.} \bibnamefont{Belik}}
\bibinfo{journal}{\prb} \textbf{\bibinfo{volume}{86}},
\bibinfo{pages}{054112} (\bibinfo{year}{2012}).

\bibitem[{\citenamefont{Baron et al.}(2000)}]{Baron2000}
\bibinfo{author}{\bibfnamefont{A.~Q.~R.} \bibnamefont{Baron}},
\bibinfo{author}{\bibfnamefont{Y.} \bibnamefont{Tanaka}},
\bibinfo{author}{\bibfnamefont{S.} \bibnamefont{Goto}},
\bibinfo{author}{\bibfnamefont{K.} \bibnamefont{Takeshita}},
\bibinfo{author}{\bibfnamefont{T.} \bibnamefont{Matsushita}} \bibnamefont{and}
\bibinfo{author}{\bibfnamefont{T.} \bibnamefont{Ishikawa}},
\bibinfo{journal}{J. Phys. Chem. Solids} \textbf{\bibinfo{volume}{61}},
\bibinfo{pages}{461} (\bibinfo{year}{2000}).








\bibitem[{\citenamefont{Riste et al.}(1971)}]{Riste71}
\bibinfo{author}{\bibfnamefont{T.} \bibnamefont{Riste}},
\bibinfo{author}{\bibfnamefont{E.~J.} \bibnamefont{Samuelsen}},
\bibinfo{author}{\bibfnamefont{K.} \bibnamefont{Otnes}} \bibnamefont{and}
\bibinfo{author}{\bibfnamefont{J.} \bibnamefont{Feder}},
\bibinfo{journal}{Solid State Commun.} \textbf{\bibinfo{volume}{9}},
\bibinfo{pages}{1455} (\bibinfo{year}{1971}).

\bibitem[{\citenamefont{Shapiro et al.}(1972)}]{Shapiro72}
\bibinfo{author}{\bibfnamefont{S.~M.} \bibnamefont{Shapiro}},
\bibinfo{author}{\bibfnamefont{J.~D.} \bibnamefont{Axe}},
\bibinfo{author}{\bibfnamefont{G.} \bibnamefont{Shirane}} \bibnamefont{and}
\bibinfo{author}{\bibfnamefont{T.} \bibnamefont{Riste}},
\bibinfo{journal}{\prb} \textbf{\bibinfo{volume}{6}},
\bibinfo{pages}{4332} (\bibinfo{year}{1972}).

\bibitem[{\citenamefont{Muller and Berlinger}(1971)}]{Muller71}
\bibinfo{author}{\bibfnamefont{K.~A.} \bibnamefont{M$\rm \ddot{u}$ller}} \bibnamefont{and}
\bibinfo{author}{\bibfnamefont{W.} \bibnamefont{Berlinger}},
\bibinfo{journal}{\prl} \textbf{\bibinfo{volume}{26}},
\bibinfo{pages}{13} (\bibinfo{year}{1971})

\bibitem[{\citenamefont{Dick and Overhauser}(1958)}]{Dick58}
\bibinfo{author}{\bibfnamefont{B.~G.} \bibnamefont{Dick,~JR.}} \bibnamefont{and}
\bibinfo{author}{\bibfnamefont{A.~W.} \bibnamefont{Overhauser}},
\bibinfo{journal}{Phys. Rev.} \textbf{\bibinfo{volume}{112}},
\bibinfo{pages}{90} (\bibinfo{year}{1958}).

\bibitem[{\citenamefont{Openphonon}()}]{Openphonon}
\bibinfo{author}{\bibfnamefont{A.} \bibnamefont{Mirone}} \bibnamefont{and}
\bibinfo{author}{\bibfnamefont{M.} \bibnamefont{d'Astuto}},
\bibinfo{journal}{available at http://sourceforge.net/projects/openphonon}.

\bibitem[{\citenamefont{Stirling}(1972)}]{Stirling72}
\bibinfo{author}{\bibfnamefont{W.~G.} \bibnamefont{Stirling}},
\bibinfo{journal}{J. Phys. C: Solid State Phys.} \textbf{\bibinfo{volume}{5}},
\bibinfo{pages}{2711} (\bibinfo{year}{1972}).

\bibitem[{\citenamefont{Cowley}(1964)}]{Cowley64}
\bibinfo{author}{\bibfnamefont{R.~A.} \bibnamefont{Cowley}},
\bibinfo{journal}{Phys. Rev.} \textbf{\bibinfo{volume}{134}},
\bibinfo{pages}{A981} (\bibinfo{year}{1964}).

\bibitem[{\citenamefont{Harada}(1970)}]{Harada70}
\bibinfo{author}{\bibfnamefont{J.} \bibnamefont{Harada}},
\bibinfo{author}{\bibfnamefont{J.~D.} \bibnamefont{Axe}} \bibnamefont{and}
\bibinfo{author}{\bibfnamefont{G.} \bibnamefont{Shirane}},
\bibinfo{journal}{Acta Cryst.} \textbf{\bibinfo{volume}{A26}},
\bibinfo{pages}{608} (\bibinfo{year}{1970}).




\bibitem[{\citenamefont{Kamba et al}(2012)}]{Kamba12}
\bibinfo{author}{\bibfnamefont{S.} \bibnamefont{Kamba}},
\bibinfo{author}{\bibfnamefont{V.} \bibnamefont{Goian}},
\bibinfo{author}{\bibfnamefont{M.} \bibnamefont{Orlita}},
\bibinfo{author}{\bibfnamefont{D.} \bibnamefont{Nuzhnyy}},
\bibinfo{author}{\bibfnamefont{J.~H.} \bibnamefont{Lee}},
\bibinfo{author}{\bibfnamefont{D.~G.} \bibnamefont{Schlom}},
\bibinfo{author}{\bibfnamefont{K.~Z.} \bibnamefont{Rushchanskii}},
\bibinfo{author}{\bibfnamefont{M.} \bibnamefont{Le\v{z}ai\'{c}}},
\bibinfo{author}{\bibfnamefont{T.} \bibnamefont{Birol}},
\bibinfo{author}{\bibfnamefont{C.~J.} \bibnamefont{Fennie}},
\bibinfo{author}{\bibfnamefont{P.} \bibnamefont{Gemeiner}},
\bibinfo{author}{\bibfnamefont{B.} \bibnamefont{Dkhil}},
\bibinfo{author}{\bibfnamefont{V.} \bibnamefont{Bovtun}},
\bibinfo{author}{\bibfnamefont{M.} \bibnamefont{Kempa}},
\bibinfo{author}{\bibfnamefont{J.} \bibnamefont{Hlinka}} \bibnamefont{and}
\bibinfo{author}{\bibfnamefont{J.} \bibnamefont{Petzelt}},
\bibinfo{journal}{\prb} \textbf{\bibinfo{volume}{85}},
\bibinfo{pages}{094435} (\bibinfo{year}{2012}).


\bibitem[{\citenamefont{Kim et al}(2012)}]{Kim12}
\bibinfo{author}{\bibfnamefont{J-W} \bibnamefont{Kim}},
\bibinfo{author}{\bibfnamefont{P.} \bibnamefont{Thompson}},
\bibinfo{author}{\bibfnamefont{S.} \bibnamefont{Brown}},
bibinfo{author}{\bibfnamefont{P.~S.} \bibnamefont{Normile}},
\bibinfo{author}{\bibfnamefont{J.~A.} \bibnamefont{Schlueter}},
\bibinfo{author}{\bibfnamefont{A.} \bibnamefont{Shkabko}},
\bibinfo{author}{\bibfnamefont{A.} \bibnamefont{Weidenkaff}} \bibnamefont{and}
\bibinfo{author}{\bibfnamefont{P.~J.} \bibnamefont{Ryan}},
\bibinfo{journal}{arXiv:1206.5417v1}







\end{thebibliography}
\end{document}